\documentclass[reprint,onecolumn,a4paper,aip,jcp,showpacs,floatfix,nofootinbib]{revtex4-1}
\usepackage{dcolumn}

\usepackage{amsmath}
\usepackage{amsfonts}

\newcommand{\dij}{\delta^{ij}}

\newcommand{\braket}[3]{\left\langle#1\left|\rule{0pt}{11pt} #2\right|#3\right\rangle}

\newcommand{\nXs}{{\nabla}_{\!X}^2}

\newcommand{\oh}{\frac{1}{2}}

\newcommand{\cross}{\times}

\newcommand{\br}{{\vec{r}}}
\newcommand{\bR}{{\vec{R}}}
\newcommand{\bn}{{\vec{n}}}

\newcommand{\riA}{r_{1A}}
\newcommand{\rjA}{r_{2A}}
\newcommand{\riB}{r_{1B}}
\newcommand{\rjB}{r_{2B}}
\newcommand{\rij}{r_{12}}

\newcommand{\harm}{Y^J_M}

\newcommand{\icm}{\mathrm{cm}^{-1}}

\newcommand*{\cent}[1]{\multicolumn{1}{c}{$#1$}}
\newcolumntype{w}[1]{D{.}{.}{#1}}

\newcommand{\HS}{H_{\Sigma\Sigma}}
\newcommand{\ES}{E_\Sigma}
\newcommand{\dH}{H}
\newcommand{\dHPS}{\mathbb{H}_{\Pi\Sigma}}
\newcommand{\dHSP}{\mathbb{H}_{\Sigma\Pi}}
\newcommand{\dHPD}{\mathbb{H}_{\Pi\Delta}}
\newcommand{\dHDP}{\mathbb{H}_{\Delta\Pi}}
\newcommand{\yP}{\frac{1}{\ES-H_{\Pi\Pi}}\,\dH_{\Pi\Sigma}\,\Phi_\Sigma^{(0)}}

\newcommand{\sot}{(\rho^i {\rho'}^j)^{(2)}}

\newcommand{\SE}{Schr{\"o}dinger equation}

\newcommand{\add}[1]{{#1}}

\begin{document}
\title{Nonadiabatic rotational states of the hydrogen molecule}

\author{Krzysztof Pachucki}
\affiliation{Faculty of Physics, University of Warsaw, Pasteura 5, 02-093 Warsaw, Poland}

\author{Jacek Komasa}
\affiliation{Faculty of Chemistry, Adam Mickiewicz University, Umultowska 89b, 61-614 Pozna{\'n}, Poland}

\date{\today}

\begin{abstract}
We present a new computational method for the determination
of energy levels in four-particle systems like H$_2$, HD, and HeH$^+$
using explicitly correlated exponential basis functions and analytic integration formulas.
In solving the Schr{\"o}dinger equation, no adiabatic 
separation of the nuclear and electronic degrees of freedom is introduced.
We provide formulas for the coupling between the rotational and electronic 
angular momenta, which enable calculations of arbitrary rotationally
excited energy levels. To illustrate the high numerical efficiency of the method, 
we present results for various states of the hydrogen molecule.
The relative accuracy to which we determined the nonrelativistic energy
reached the level of $10^{-12}$--$10^{-13}$, which corresponds to an uncertainty 
of $10^{-7}$--$10^{-8}\,\icm$.
\end{abstract}

\maketitle
\section*{}
\vspace{-1cm}

\section{Introduction}

The hydrogen molecule gives an opportunity to test the foundations of quantum chemistry,
which are based on quantum electrodynamic theory. In principle, there are no 
limits to the theoretical precision of the determination of molecular levels, apart from
the limited accuracy of fundamental constants, such as the electron-proton mass ratio,
the Rydberg constant, or the nuclear mean square charge radii. This opens up the possibility
to determine these fundamental constants from molecular spectroscopic data, or alternatively
to look for any discrepancies between theoretical predictions and experimental results
to search for as yet unknown interactions.\cite{Ubachs:16} In fact, in recent years
we have observed significant progress in the accuracy of molecular spectroscopy.
\cite{Liu:09,Liu:10,Sprecher:10,Sprecher:11,Kassi:12b,Sprecher:13,Niu:14,Niu:15,Wcislo:16}
In the particular case of the hydrogen molecule,
contemporary measurements have reached the accuracy of $10^{-5}\,\icm$ (relative $10^{-9}$) for
selected transitions.\cite{Cheng:12,Mondelain:16,Biesheuvel:16}
On the theoretical side, various relativistic and quantum electrodynamic corrections
have recently been calculated,\cite{PKP17,PKCP16} but the principal problem up to now
has been the insufficient accuracy of nonrelativistic energy levels.
In a general multiparticle case, the complexity of the \SE\ prevents its accurate solution
and enforces approximations to be made. The most common one is the adiabatic approximation,
which assumes the separation of the electronic and nuclear dynamics.
Only a few attempts to solve directly, {\em i.e.} without the adiabatic approximation,
the four-body \SE\ for H$_2$ have been published.
The first successful method was developed by Ko{\l}os and Wolniewicz 
over 50 years ago.\cite{Kolos:63,Kolos:64b} 
They employed a nonadiabatic expansion of a trial wave function 
in products of electronic James-Coolidge basis functions\cite{James:33} 
and the vibrational functions of the form
$h_n(R)=R^{-3}\,e^{-x^2/2}\,\mathcal{H}_n(x)$
with $x=\beta\,|R-R_\mathrm{e}|$, where $\beta$ and $R_\mathrm{e}$ are variational parameters,
and $\mathcal{H}_n$ denotes the $n$-th Hermite polynomial. The expansion was composed of 54 electronic 
terms and six $h_n$ functions yielding 147-terms in total. Because of this relatively short expansion, 
the obtained nonrelativistic dissociation energy $D_0=36\,114.7\,\icm$ differed by ca. $3\,\icm$ 
from the exact value. Nevertheless, the pioneering work by Ko{\l}os and Wolniewicz has set
the foundations of the theoretical techniques for accurate calculations and, regarding the then 
available computing capabilities, should be considered as a great success of theory.
Fifteen years later, using the same type of wave functions with significantly larger
(1070-term) and carefully optimized expansion, a refined integration method, and much more
powerful computers, Bishop and Cheung\cite{Bishop:78} reduced the error to $0.2\,\icm$.
Quite a different approach, based on the quantum Monte Carlo method, was presented by
Traynor et al.\cite{Traynor:91} in 1991 and improved later by
Chen and Anderson\cite{Chen:95}. Their trial wave function was a product
of four terms $\psi_i$. The first two terms were a combination of one-electron functions
centered on nuclei $A$ and $B$,
$\psi_i=e^{-a\,r_{iA}}+e^{-a\,r_{iB}}$\,.
The third term was the Jastrow factor responsible for interparticle correlation and
cusps
$\psi_3=\exp\left(\sum_{ij}\frac{a_{ij}r_{ij}}{1+b_{ij}r_{ij}}\right)$
and the last term accounted for nuclear vibration and was of the Gaussian form
$\psi_4=e^{-d(R-c)^2}$.
The quantum Monte Carlo method allowed the finite basis set error to be eliminated
but introduced instead a statistical (sampling) error, which in the latter
calculations was of about $\pm 0.2\,\icm$.
A breakthrough result has been published by Kinghorn and Adamowicz.\cite{Kinghorn:99,Kinghorn:00}
Using a 512-term basis of explicitly correlated Gaussian functions, they have diminished
the error in the nonrelativistic $D_0$ to $1.7\cdot 10^{-3}\,\icm$. Later on,
successively improving the optimization technique and expanding the basis set size to $10\,000$ terms,
Bubin and Adamowicz\cite{Bubin:03,Bubin:09b,Mitroy:13} arrived at an extremely accurate
solution of the four-particle \SE\ to obtain $D_0=36\,118.797\,74(1)\,\icm$.

All these calculations have been limited to the nonrotational state of the molecule ($J=0$).
In this work, we show how to incorporate the coupling between the rotational
and electronic angular momentum in the straightforward manner and
increase the accuracy of the nonrelativistic dissociation energy
up to the level of $10^{-7}$--$10^{-8}\,\icm$
for the ground as well as for the rotationally and vibrationally excited energy levels
of the electronic X${}^1\Sigma_\mathrm{g}^+$ state. This project comprises one of the stages
heading toward the prediction of the total energies
of the hydrogen molecule with the accuracy of $10^{-6}\,\icm$. The other contributions
are relativistic $O(\alpha^2)$, leading quantum electrodynamics $O(\alpha^3)$,
and higher order $O(\alpha^4)$ which are known only within the adiabatic approximation.\cite{PKCP16}
The knowledge of nonadiabatic wave functions obtained here,
is essential for the calculation of these contributions.

\section{Method}

\subsection{From a general exponential to the nonadiabatic James-Coolidge basis function}

The method described here is relevant to a molecule consisting of two electrons, 
labeled 1 and 2, and two nuclei, labeled $A$ and $B$, with masses $M_A$ and $M_B$ and charges $Z_A$ and $Z_B$.
The nonrelativistic Hamiltonian for this system is
\begin{align}\label{EHam}
H&=-\frac{1}{2\,M_A}\nabla_{\!A}^2-\frac{1}{2\,M_B}\nabla_{\!B}^2
-\oh\nabla_1^2-\oh\nabla_2^2\nonumber\\
&\quad\ +\frac{1}{\rij}+\frac{Z_A\,Z_B}{r_{AB}}-\frac{Z_A}{r_{1A}}-\frac{Z_A}{r_{2A}}
-\frac{Z_B}{r_{1B}}-\frac{Z_B}{r_{2B}}\,.
\end{align}
We start the description of the method with a general exponential basis function of the following 
\add{translationally invariant } form
\begin{eqnarray}\label{EExpBas}
\Phi_{\{k\}}&=e^{-u_1\,R-w_1\,\rij-y\,\eta_1-x\,\eta_2-u\,\zeta_1-w\,\zeta_2}\,\nonumber\\
&\quad \cross R^{k_0}\,\rij^{k_1}\,\eta_1^{k_2}\,\eta_2^{k_3}\,\zeta_1^{k_4}\,\zeta_2^{k_5}\,,
\end{eqnarray}
where $u_1$, $w_1$, $y$, $x$, $u$, and $w$ are real numbers, whereas $k_i$ are nonnegative integers,
and where
\begin{align}
&\zeta_1=\riA+\riB, \qquad \eta_1=\riA-\riB,\nonumber\\
&\zeta_2=\rjA+\rjB, \qquad \eta_2=\rjA-\rjB, \qquad
\bR = \br_{AB}\,.\label{KPcoords}
\end{align}
By setting $u_1\equiv \alpha$, $w_1=0$, $y=0$, $x=0$, and $u=w=\beta$ 
we arrive at simplified basis functions
\begin{equation}
\Phi_{\{k\}}=e^{-\alpha\,R-\beta(\zeta_1+\zeta_2)}\,
R^{k_0}\,\rij^{k_1}\,\eta_1^{k_2}\,\eta_2^{k_3}\,\zeta_1^{k_4}\,\zeta_2^{k_5}\,,\label{EJCBas}
\end{equation}
which still form a complete basis set.
We call this function the nonadiabatic James-Coolidge (naJC) function for its resemblance to the original
James-Coolidge (JC) basis function used in fixed-nuclei calculations. The difference between our
nonadiabatic function and the JC function is in the internuclear correlation factor
as well as in the meaning of the $\zeta$ and $\eta$ variables.

\subsection{The variational nonadiabatic wave function for an arbitrary rotational angular momentum}
\label{Sec:var}

The rotational angular momentum of nuclei couples to the electronic angular momentum, $\vec{L}$,
and gives the total angular momentum $\vec{J}$ of a molecule.
For this reason, the wave function $\Psi^{J,M}$ of a rotational level $J$ (formally depending also 
on the projection of $\vec{J}$ on the $Z$ axis in the laboratory frame) must contain components 
that describe the electronic $\Sigma$, $\Pi$, $\Delta$, \dots\ states. 
In the following set of formulas we construct such a wave function  
and we introduce a necessary notation. The total wave function is a sum
of the components with growing $\Lambda$---the eigenvalue of the $\vec n\cdot\vec L$ operator
\begin{align}
\Psi^{J,M}&=\Psi^{J,M}_\Sigma+\Psi^{J,M}_\Pi+\Psi^{J,M}_\Delta+\dots\label{WFJ:total}
\intertext{where}
\Psi^{J,M}_\Sigma&=\harm\,\Phi^J_\Sigma\,,\qquad \mbox{for}\;J\geq 0 \label{WFJ:Sigma}\\
\Psi^{J,M}_\Pi&=\sqrt{\frac{2}{J(J+1)}}R\,\rho^i\left(\nabla_R^i\harm\right)\Phi^J_\Pi\,,\qquad \mbox{for}\;J\geq 1\label{WFJ:Pi}\\
\Psi^{J,M}_\Delta&=\sqrt{\frac{4}{(J-1)J(J+1)(J+2)}} 
R^{2}\,\sot\left(\nabla_{\!R}^i\nabla_{\!R}^j\harm\right)
\,\Phi^J_\Delta \qquad\mbox{for}\;  J\geq 2\,. \label{WFJ:Delta}
\end{align}
The particular form of functions in Eqs. (\ref{WFJ:Sigma})--(\ref{WFJ:Delta})
is convenient for the calculation of matrix elements as for example, the overlap matrix is block diagonal.
In the above equations we use the following notation 
\begin{align}
\sot&\equiv\oh\Bigl(\rho^i\rho'^j+\rho^j\rho'^i-\left(\dij-n^in^j\right)\,\vec\rho\cdot\vec\rho\,'\Bigr)\\
\vec\rho,\vec\rho\,'&\equiv\vec\rho_1\text{ or }\vec\rho_2\\  
\rho_a^i&=\left(\dij-n^in^j\right)\,r_{aB}^j =\left(\dij-n^in^j\right)\,r_{aA}^j
 \nonumber\\&\qquad \quad \mbox{with}\quad n^i=\frac{R^i}{R}\,
\end{align}
and we assume the Einstein summation convention, i.e. an implicit sum over all values of
a repeated Cartesian index. 
The symbol $\harm=\harm(\bn)$ denotes the spherical harmonic.
The functions $\Phi^J_{\Lambda}$ represent
linear expansions in the above-defined naJC basis functions~(\ref{EJCBas})
\begin{equation}\label{Ele}
\Phi^J_{\Lambda}= R^J\,\sum_{\{k\}} c_{\{k\}} \left(1+\mathcal{P}_{12}\right)\,\Phi^J_{{\Lambda}\{k\}}
\end{equation}
for ${\Lambda}=\Sigma,\Pi,\Delta,\dots\,$. For each pair $J$ and ${\Lambda}$, 
the function $\Phi^J_{\Lambda}$ has its own set of nonlinear parameters, 
therefore we distinguish $\Phi_{\{k\}}$ of Eq. (\ref{EJCBas}) by indices $J$ and ${\Lambda}$. 
\add{In the equation above, the symbol $\mathcal{P}_{12}$ means the electron permutation operator
and the linear coefficients $c_{\{k\}}$ are determined variationally.}

\add{
The nuclear rotation in the wave function $\Psi^{J,M}$ is described 
by the spherical harmonics $\harm(\bn)$, whereas the electronic angular contribution
is represented in the form of the expansion~(\ref{WFJ:total}) in Cartesian coordinates $\rho^i$.
Each term of this expansion represents a function with a well-defined projection
of the electronic angular momentum $\Lambda$.
Moreover, the product of $\vec\rho\cdot\vec\nabla_R\harm$ commutes with the total
angular momentum operator $\vec J$  so that it preserves the correct $J$ and $M$ quantum numbers.
Finally, this expansion is complete. In practice, though, it can be cut 
due to the rapidly decreasing contribution from the subsequent terms.
}

A note concerning a linear dependence and a completeness of the basis set is in place here.
To ensure the completeness, the function $\Psi^{J,M}_\Delta$ appears in two variants. 
The one in which both $\rho$ and $\rho'$ point at the same electron, and the other, 
in which they point at different electrons.
Certain combinations of these two variants \add{are linearly dependent},
which originates from the following identity
\begin{align}
2\,\vec\rho_1\vec\rho_2\,(\rho_1^i\,\rho_2^j)^{(2)}
&=\vec\rho_2^{\;2}\,(\rho_1^i\,\rho_1^j)^{(2)}+\vec\rho_1^{\;2}\,(\rho_2^i\rho_2^j)^{(2)}.
\end{align}
This linear dependence can be avoided by, for example, 
\add{using } the second variant basis functions with at least one
$k_i=0$, for $i=2,\dots,5$.

\subsection{Symmetry of the wave function}

The nonrelativistic Hamiltonian~(\ref{EHam}) is invariant under translation, rotation, 
and spatial inversion $\hat{P}$. The inversion $\hat{P}$
reverses the sign of spatial coordinates of all particles leaving their spin
unchanged, and the wave function is an eigenstate of $\hat{P}$ with eigenvalues $\pm 1$.
The wave function has also a definite symmetry with respect to the exchange of electrons 
---it is either symmetric or antisymmetric for the total electronic spin $S=0$ or 1, respectively.
In practice, this symmetry is enforced by acting on the spatial wave function
with the $\oh(1\pm \hat{P}_{12})$ operator, where $\hat{P}_{12}$ exchanges the electron labels. 

For a homonuclear molecule, additional symmetries arise.
The {\em gerade/ungerade} inversion symmetry is the inversion of all electronic coordinates
with respect to the geometric center of a molecule.
Recalling how the inversion operator $\hat{\imath}$ acts on the electronic variables
$\hat{\imath}\,\zeta_i=\zeta_i$, $\hat{\imath}\,\eta_i=-\eta_i$, and $\hat{\imath}\,\rij=\rij$, 
one finds that $\hat{\imath}\,\Phi_{\{k\}}=(-1)^{k_2+k_3}\Phi_{\{k\}}$ and hence
\begin{equation}
\hat{\imath}\,\Phi_{\{k\}}=
\begin{cases}
+\Phi_{\{k\}}\,, & \text{for $k_2+k_3$ even ({\em gerade}}) \\
-\Phi_{\{k\}}\,, & \text{for $k_2+k_3$ odd ({\em ungerade}}) \,.\\
\end{cases}
\end{equation}
A wave function of a homonuclear molecule also has a symmetry due to the exchange 
of the nuclei. For a specified {\em gerade/ungerade} inversion symmetry 
and the total nuclear spin, only even or odd $J$ levels are allowed
depending on the statistics (boson or fermion) of the nuclei. For example,
if we restrict our considerations to the electronic ground state (${}^1\mathrm{\Sigma}_g^+$) of H$_2$, 
we observe that the subsequent rotational levels assume alternate nuclear spins.
As a result, the even $J$ levels correspond to the nuclear singlet ({\em para}-hydrogen)
and the odd $J$ levels to the triplet state ({\em ortho}-hydrogen).

\subsection{Reduction of the angular factor}

An important step in the analytic evaluation of the matrix elements with functions $\Psi^{J,M}_{\Lambda}$
is the integration over the nuclear angular variables.
In this section, we supply formulas for the reduction of the general matrix elements
by performing the integration $\int d\Omega_R$. Let us first note, that
for an arbitrary scalar operator $Q$ we have
$\braket{\Psi^{J,M}_\Lambda}{Q}{\Psi^{J,M'}_\Lambda}\sim\delta_{M,M'}$.
In the simplest case of
matrix elements with a scalar electronic ({\em i.e.} containing no differentiation over $R$) 
operator $Q_{\mathrm el}$
we have
\begin{align}
\braket{\Psi^{J,M}_\Sigma}{Q_{\mathrm el}}{\Psi^{J,M}_\Sigma}
  &=\braket{\Phi_\Sigma^J}{Q_{\mathrm el}}{\Phi_\Sigma^J}\\
%
\braket{\Psi^{J,M}_\Pi}{Q_{\mathrm el}}{\Psi^{J,M}_\Pi}
	&=\braket{\rho^i\,\Phi_\Pi^J}{Q_{\mathrm el}}{\rho^i\,\Phi_\Pi^J}\\
\braket{\Psi^{J,M}_\Delta}{Q_{\mathrm el}}{\Psi^{J,M}_\Delta}
	&=\braket{\sot\,\Phi_\Delta^J}{Q_{\mathrm el}}{\sot\,\Phi_\Delta^J}
\end{align}
and all the off-diagonal matrix elements vanish. The next set of formulas applies to the
diagonal matrix elements with the nuclear kinetic energy operator

\begin{equation}
\hat{T}=-\frac{1}{2\,M_A}\nabla_{\!A}^2-\frac{1}{2\,M_B}\nabla_{\!B}^2\,,
\end{equation}
namely
\begin{align}
\braket{\Psi^{J,M}_\Sigma}{\nXs}{\Psi^{J,M}_\Sigma}
  &=\braket{\Phi_\Sigma^J}{\nXs-J(J+1)\,R^{-2}}{\Phi_\Sigma^J}\\[2ex]
%
\braket{\Psi^{J,M}_\Pi}{\nXs}{\Psi^{J,M}_\Pi}&=
\braket{\rho^i\,\Phi_\Pi^J}{\nXs-[J(J+1)-2]\,R^{-2}}{\rho^i\,\Phi_\Pi^J}\\[2ex]
%
\braket{\Psi^{J,M}_\Delta}{\nXs}{\Psi^{J,M}_\Delta}  &=
\braket{\sot\,\Phi_\Delta^J}{\nXs-[J(J+1)-6]\,R^{-2}}{\sot\,\Phi_\Delta^J}
	\intertext{with $X=A$ or $B$. Finally, the nondiagonal matrix elements read}
%
\braket{\Psi^{J,M}_\Pi}{\nXs}{\Psi^{J,M}_\Sigma}
  &=\pm\sqrt{2\,J(J+1)}\braket{\rho^i\,\Phi_\Pi^J}{R^{-1}}{\nabla_{\!X}^i\Phi_\Sigma^J}\\
%
\braket{\Psi^{J,M}_\Delta}{\nXs}{\Psi^{J,M}_\Pi}  &=
  \pm\sqrt{2\,(J-1)(J+2)}\braket{\sot\,\Phi_\Delta^J}{R^{-1}}{\rho^j\nabla_{\!X}^i\Phi_\Pi^J}.
\end{align}
where $+$ and $-$ is for $X=A$ and $B$, correspondingly, with $\vec R = \vec R_A - \vec R_B$.
All the remaining matrix elements vanish, so that 
the overlap $\mathbb{N}$ and Hamiltonian $\mathbb{H}$ matrices have the following block-band structure
\begin{equation}\label{NHmatrices}
\mathbb{N}=
\begin{pmatrix}
\mathbb{N}_{\Sigma\Sigma} & 0 & 0 & \cdots \\
0 & \mathbb{N}_{\Pi\Pi} & 0 & \cdots \\
0 & 0 & \mathbb{N}_{\Delta\Delta} & \cdots \\
\vdots & \vdots & \vdots & \ddots
\end{pmatrix}
\qquad \mathbb{H}=
\begin{pmatrix}
\mathbb{H}_{\Sigma\Sigma} & \dHSP & 0 & \cdots \\
\dHPS & \mathbb{H}_{\Pi\Pi} & \dHPD & \cdots \\
0 & \dHDP & \mathbb{H}_{\Delta\Delta} & \cdots \\
\vdots & \vdots & \vdots & \ddots
\end{pmatrix}.
\end{equation}


\subsection{Integrals with the exponential function}

The previous section dealt with matrix elements without any reference to a specific
shape of the spatial part of the basis function. Therefore, the above formulas can be utilized
also with types of basis functions other than that presented in this article, e.g. 
with explicitly correlated Gaussian functions.\cite{Mitroy:13}
The present section, in turn, is devoted to the exponential basis functions,
in particular to the naJC functions of Eq.~(\ref{EJCBas}). We start, however, with the most general
integral for a four-body system
\begin{align}\label{IGeneral}
  G&=\int \frac{d^3 R}{4\,\pi}\,\int \frac{d^3 r_{1A}}{4\,\pi}\,\int \frac{d^3 r_{2A}}{4\,\pi}\,
  e^{-u_1\,R-w_1\,\rij-y\,\eta_1-x\,\eta_2-u\,\zeta_1-w\,\zeta_2}\,
R^{n_0}\,\rij^{n_1}\,\eta_1^{n_2}\,\eta_2^{n_3}\,\zeta_1^{n_4}\,\zeta_2^{n_5}/\cal{D}
\end{align}
where
\begin{equation}
  {\cal D} = R\,r_{12}\,r_{1A}\,r_{1B}\,r_{2A}\,r_{2B} =
  \frac{1}{16}\,R\,\rij\,(\zeta_1+\eta_1)\,(\zeta_1-\eta_1)\,(\zeta_2+\eta_2)\,(\zeta_2-\eta_2)\,.
\end{equation}
The matrix elements of the Hamiltonian evaluated in the exponential basis~(\ref{EExpBas}) 
can be expressed by a combination of integrals belonging to the class defined in Eq.~(\ref{IGeneral}). 
All of these integrals can be obtained through a recurrence relation 
starting from the so called master integral
\begin{align}\label{Intg}
  g&=\int \frac{d^3 R}{4\,\pi}\,\int \frac{d^3 r_{1A}}{4\,\pi}\,\int \frac{d^3 r_{2A}}{4\,\pi}\,
  \frac{e^{-u_1\,R-w_1\,\rij-y\,\eta_1-x\,\eta_2-u\,\zeta_1-w\,\zeta_2}}{\cal{D}}\,.
\end{align}
The analytical form of the integral~(\ref{Intg}) was obtained
by Fromm and Hill\cite{Fromm:87} in 1987. Their result, although terribly 
troublesome for a numerical evaluation, was a milestone in the evaluation
of the four-body exponential integrals. A special case of this integral 
was evaluated analytically by Remiddi\cite{Remiddi:91}, who expressed his result in terms
of the logarithmic and the Euler dilogarithmic functions. 
In 1997 a significant simplification of the result
obtained by Fromm and Hill was achieved by Harris\cite{Harris:97}, who managed to eliminate the original
singularities and arrive to a much more computationally friendly formulation.
Another significant step in this field was made in 2009 when the effective recurrence relations
were discovered\cite{Pachucki:09} enabling evaluation of an arbitrary integral out of the whole 
class given by Eq.~(\ref{IGeneral}).

The master integral $g$ and its derivatives satisfy the following differential equations\cite{Pachucki:09}
\begin{equation}
\sigma\,\frac{\partial g}{\partial a } +
\frac{1}{2}\,\frac{\partial\sigma}{\partial a}\,g + P_a = 0\,, \label{EME}
\end{equation}
where $a$ is one of the parameters $u_1\equiv t$, $w_1$, $y$, $x$, $u$, or $w$, and where
\begin{eqnarray}
\sigma &=&\sigma_0 +t^2\,\sigma_2 + t^4\,\sigma_4, \label{Esigma}\\
\sigma_0 &=& w_1^2\,(u + w - x - y)\,(u - w + x - y)\,(u - w - x + y)\,(u + w + x + y) \nonumber \\&&
        + 16\,(w\,x - u\,y)\,(u\,x - w\,y)\,(u\,w - x\,y)\,,\nonumber \\
\sigma_2 &=& w_1^4-2\,w_1^2\,(u^2+w^2+x^2+y^2)+16\,u\,w\,x\,y\,,\nonumber \\
\sigma_4 &=& w_1^2\,.\nonumber
\end{eqnarray}
The inhomogeneous term $P_a$ is a combination of several logarithmic functions
and is presented explicitly in Appendix A of Ref.~\citenum{Pachucki:12b}.

The most general integral of Eq.~(\ref{IGeneral}) can be obtained by successive, multiple differentiation of
the master integral $g$
\begin{align}\label{IntG}
&G(n_0,n_1,n_2,n_3,n_4,n_5)\nonumber\\
&=\left(-\frac{\partial}{\partial t}\right)^{n_0}
\left(-\frac{\partial}{\partial w_1}\right)^{n_1}
\left(-\frac{\partial}{\partial y}\right)^{n_2}
\left(-\frac{\partial}{\partial x}\right)^{n_3}
\left(-\frac{\partial}{\partial u}\right)^{n_4}
\left(-\frac{\partial}{\partial w}\right)^{n_5}\,
g(t,w_1,y,x,u,w).
\end{align}
Each differentiation raises by one the power of the associated variable in the pre-exponential factor
of the integrand in Eq.~(\ref{Intg}). 
However, from the practical point of view, it is much more convenient to use
recursion relations for raising the powers $n_i$.
Let us briefly overview the steps leading to these recurrence relations.
First, in Eq.~(\ref{EME}) we set $a=y$ and generate the pertinent inhomogeneous term $P_y$.
Next, we differentiate Eq.~(\ref{EME}) $n_2$ times with respect to $y$ and then set $y=0$.
We proceed analogously with the variables $x$ and $w_1$, and obtain the relation connecting
different integrals of the $G$-class. From this equation we extract $G(n_0,n_1,n_2,n_3,n_4,n_5)$
and obtain the recurrence relation which, starting from $G(0,0,0,0,0,0)$, enables  
the integrals $G$ to be obtained for an arbitrary combination of non-negative integers $n_i$,
expressed in terms of derivatives of $P_y$. The multiple derivatives of $P_y$ are combinations 
of rational and logarithmic functions, and are numerically stable for $t-2u$ sufficiently far from zero. 
This condition can be easily satisfied and does not introduce limitations in practical calculations.
Setting $w_1$, $y$, and $x$ to zero simplifies significantly the analytic expressions for integrals 
in the naJC basis, in particular the $\sigma$ from Eq.~(\ref{Esigma}) vanishes. 

A small sample of explicit expressions for $G$ and $P_y$ is given below (for $w=u$).
Note that the master integral~(\ref{Intg}) for the naJC basis is represented explicitly by $G(0,0,0,0,0,0)$.
\begin{align}
G(0,0,0,0,0,0)&= P_y(0,0,1,0,0,0)/(16 u^4) \\
G(0,0,0,0,0,1)&= P_y(0,0,1,0,0,0)/(8 u^5)+P_y(0,0,1,0,0,1)/(16 u^4) \\
G(0,0,0,0,1,0)&= G(0,0,0,0,0,1) \\
G(0,0,0,1,0,0)&= 0 \\
G(0,0,1,0,0,0)&= 0 \\
G(0,1,0,0,0,0)&= P_y(0,1,1,0,0,0)/(16 u^4) 
\end{align}
where
\begin{align}
P_y(0,0,1,0,0,0)&= \frac{-16 u^3 \ln{(2u)}}{t(t+2u)}+\frac{16u^3\ln{(4u)}}{(t-2u)(t+2u)}
 -\frac{32u^4\ln{(t+2u)}}{t(t - 2 u)(t + 2 u)} \\
P_y(0,0,1,0,0,1)&=-\frac{16u^3}{(t-2u)(t+2u)^2}+\frac{16u^2(t+u)^2\ln{(2u)}}{t^2(t+2u)^2} 
 +\frac{8 u^2 (t^2 - 2 u^2) \ln{(2 u)}}{t^2(t+2u)^2}\nonumber\\&\phantom{=}\
 -\frac{8u(3t^2u-4u^3)\ln{(4u)}}{(t-2u)^2(t+2u)^2}
 +\frac{64u^3(t^2-2u^2)\ln{(t+2u)}}{t(t-2u)^2(t+2u)^2} \\
P_y(0,1, 1, 0, 0, 0)&=\frac{4 u^2}{(t + 2 u)^2} 
\end{align}

We note that, by symmetry, the integrals $G$ with $n_2+n_3$ odd vanish, as do $P_y$ with $n_2+n_3$ even.
The procedure sketched above allows the whole $G$-class of integrals
to be evaluated analytically in a simple form.

\section{Numerical approach}

\subsection{A perturbative solution of the eigenvalue problem}
\label{Sec:pert}

The solution of the \SE\ in terms of the basis functions~(\ref{EJCBas}) is now reformulated
into the generalized eigenvalue problem with the Hamiltonian $\mathbb{H}$ and
overlap $\mathbb{N}$ matrices 
\begin{equation}\label{EGSEP}
(\mathbb{H}-E\,\mathbb{N})\,\mathbb{C}=0\,,
\end{equation}
where $\mathbb{C}$ is a vector of linear coefficients from Eq.~(\ref{Ele}).
For all states this equation can be solved directly, e.g. by the inverse iteration method.
However, due to the large size of the matrices $\mathbb{H}$ and $\mathbb{N}$ for rotational
states, it is more economical to apply the inverse iteration method only for the $\Sigma$ component,
and obtain the $\Pi$ and $\Delta$ components from the standard perturbation theory.
In other words, when $J>0$, the wavefunction $\Psi^{J,M}$, as defined in Sec.~\ref{Sec:var}, 
is composed of mutually orthogonal ${\Lambda}$-segments. 
The orthogonality is manifested in the block-diagonal structure 
of the overlap matrix, Eq.~(\ref{NHmatrices}). The \add{small block-off-diagonal terms 
of $\mathbb{H}$ enable } rapidly converging perturbative expansion. In this section, we supply explicit 
formulas for the subsequent perturbational corrections.

Let us first consider the approximated energy level $E^{(0)}=E_\Sigma$ obtained from 
the unperturbed wavefunction $\Psi^{J,M}=\Psi^{J,M}_\Sigma$. 
The Rayleigh-Schr{\"o}dinger perturbation 
theory yields the second order (with respect to the power of the off-diagonal parts of the Hamiltonian)
energy shift $E_\Pi^{(2)}$
\begin{align}
E_\Pi^{(2)}&
=\braket{\Psi^{J,M}_\Sigma}{H_{\Sigma\Pi}\frac{1}{\ES-H_{\Pi\Pi}}H_{\Pi\Sigma}}{\Psi^{J,M}_\Sigma}
=\braket{\Psi^{J,M}_\Sigma}{V_\Pi(E_\Sigma)}{\Psi^{J,M}_\Sigma}
\end{align}
where 
\begin{equation}
V_\Pi(E)=H_{\Sigma\Pi}\frac{1}{E-H_{\Pi\Pi}}H_{\Pi\Sigma}.
\end{equation}
The fourth order correction $E_\Pi^{(4)} + E_\Delta^{(4)}$ can be evaluated from
\begin{align}
E_\Pi^{(4)}&=\braket{\Psi^{J,M}_\Sigma}{V_\Pi(\ES)\frac{1}{(\ES-\HS)^{'}}V_\Pi(\ES)}{\Psi^{J,M}_\Sigma}
\nonumber\\&\phantom{=}\ 
 +\braket{\Psi^{J,M}_\Sigma}{\left.\frac{\partial V_\Pi}{\partial E}\right|_{\ES}}{\Psi^{J,M}_\Sigma} E_\Pi^{(2)},
 \label{EEPis}\\
E_\Delta^{(4)}&=\braket{\Psi^{J,M}_\Sigma}{H_{\Sigma\Pi}\frac{1}{E_{\Sigma}-H_{\Pi\Pi}}H_{\Pi\Delta}\frac{1}{E_{\Sigma}-H_{\Delta\Delta}}H_{\Delta\Pi}\frac{1}{E_{\Sigma}-H_{\Pi\Pi}}H_{\Pi\Sigma}}{\Psi^{J,M}_\Sigma}.
\end{align}
Each $H_{{\Lambda}'{\Lambda}}$ Hamiltonian contains the $m/\mu$ factor of the order of $10^{-3}$, 
which makes the perturbation series converge very rapidly. In particular, the $E_\Delta^{(6)}$
correction would be approximately 5-6 orders of magnitude smaller than $E_\Delta^{(4)}$.
Therefore, taking into account just the first four terms of the perturbative expansion is sufficient for our purposes
\begin{equation}
E\approx E_\Sigma+E_\Pi^{(2)}+E_\Pi^{(4)} + E_\Delta^{(4)}.
\end{equation}

A similar perturbation expansion holds for the wave function, namely
\begin{align}\label{EPWF}
  \Phi&=
\left[\renewcommand*{\arraystretch}{1.5}
\begin{array}{c}
\Phi_\Sigma^{(0)}+\Phi_\Sigma^{(2)}+\Phi_\Sigma^{(4)}+\dots \\
\Phi_\Pi^{(1)}+\Phi_\Pi^{(3)}+\dots \\
\Phi_\Delta^{(2)}+\dots
\end{array}\right]
\end{align}
where
\begin{align}
\Phi_\Sigma^{(0)}& \ \text{is the unperturbed function} \\
\Phi_\Pi^{(1)}&=\yP\,, \\
\Phi_\Sigma^{(2)}&=\frac{1}{(\ES-\HS)'}\,\dH_{\Sigma\Pi}\,\Phi_\Pi^{(1)}
 -\oh\Bigl\langle\Phi_\Pi^{(1)}|\Phi_\Pi^{(1)}\Bigr\rangle\,\Phi_\Sigma^{(0)}\\
\Phi_\Delta^{(2)}&=\frac{1}{\ES-H_{\Delta\Delta}}\,\dH_{\Delta\Pi}\,\Phi_\Pi^{(1)}\,.
\end{align}

\subsection{Technical details}

Each naJC basis function $\Phi_{\{k\}}$, apart from the set of integers $k_i$, depends on 
two real positive parameters $\alpha_k$ and $\beta_k$. The set of the basis functions with 
a common pair $\alpha_k$ and $\beta_k$ will be called the sector.
Such a sector contains all basis functions with integer powers of $R$ ranging from $k_0^\mathrm{min}$
to $k_0^\mathrm{max}$. The total wave function can be composed of a number of such sectors.
The optimal value $k_0^\mathrm{max}$ was determined through numerical experiments 
for each state separately. The $k_0^\mathrm{min}$ in turn was set to ${\Lambda}$.
The `electronic' integer parameters $k_1\dots,k_5$ are used to organize the basis functions
in `shells'. The given basis function $\Phi_{\{k\}}$ belongs to the shell number 
$\Omega_k=\sum_{i=1}^5 k_i$. To describe a sector of basis functions 
for given $J$ and ${\Lambda}$ values,
we use a four parameter symbol $(k_0^\mathrm{max},\Omega,\alpha,\beta)$. Such a sector involves
basis functions with the nonlinear parameters $\alpha$ and $\beta$, and with the integer powers $k_i$
fulfilling $k_0^\mathrm{min}\le k_0\le k_0^\mathrm{max}$ and $0\le\Omega_k\le\Omega$.
By increasing $\Omega$ we can systematically add new basis functions to the expansion
and observe the convergence of energy with increasing total size of the basis set $K$.

To solve the eigenproblem~(\ref{EGSEP}) we employed the inverse iteration method, which consists of
the $\mathbb{M}\,\mathbb{D}\,\mathbb{M}^T$ decomposition of the $\mathbb{H}-E\,\mathbb{N}$ matrix 
followed by a solution of the linear equations set
performed several times up to the assumed convergence. 
The matrix $\mathbb{D}$ is block diagonal with blocks of the order 1 or 2, and $\mathbb{M}$
is unit lower triangular. The workload of the decomposition
step grows with the basis size like $K^3$ whereas that of the remainder steps like $K^2$; 
therefore, for large matrices the decomposition step determines the timing of all of the computations.
The linear algebra calculations, as well as the evaluation of matrix elements, were performed using extended precision arithmetics
implemented with the help of the QD library.\cite{qd} It enables nearly octuple precision (212 bit, 62 digits),
which is sufficient to obtain the required accuracy of the energy levels considered in this work.

We consider H$_2$ in its electronic ground state X$^1\Sigma_g^+$ and, 
due to limited space for tables, we restrict the presentation of numerical results
to the ground vibrational level $v=0$. This restriction, however, is not related to
limitations of the method.

\subsection{$J=0$ level}

We consider here the ground rotational level $J=0$,
which requires no coupling to the electronic states with higher angular momentum
to be involved, that is $\Psi^{J,M}=\Psi^{J,M}_\Sigma$. We used a two-sector basis 
of $\Sigma$-functions: (30,$\Omega$,$\alpha$,$\beta^{(1)}$) and (30,$\Omega$-2,$\alpha$,$\beta^{(2)}$).
The parameters $\alpha$, $\beta^{(1)}$, and $\beta^{(2)}$ were optimized variationally with respect to
the energy of the level separately for each $\Omega$. The optimal parameters, the total
size of the basis, and the resulting energy are listed in Table~\ref{Tconv}.
Extrapolation of the energy to an infinite basis set size enables determination of the recommended energy value
and its estimated numerical uncertainty. 
For this particular level, we assess the accuracy of the energy as $3\cdot 10^{-13}$ hartree. 
By subtracting the energy $E_{0,0}$ from the exactly known sum of the energy of two hydrogen atoms,
$2\,E(H)=\frac{m_p}{m_p+m_e}$ hartree, we calculated the dissociation energy $D_{0,0}$
listed in the last column of the table. The numerical accuracy of $D_{0,0}$ is estimated 
as $3\cdot 10^{-8}\,\icm$. This estimation, however, does not account for the uncertainty
originating from determination of the fundamental physical constants.
All calculations reported in this work were performed with the best currently available
values of the proton-to-electron mass ratio $m_p/m_e=1\,836.152\,673\,89(17)$ 
and of the Rydberg constant $R_\infty=109\,737.315\,685\,08(65)\,\icm$
obtained from the 2014 CODATA compilation.\cite{CODATA:14}
The uncertainties of both physical constants limit the accuracy
of our final results. On the other hand, the problem can be reversed and future
high-accuracy relativistic calculations, in connection with high-accuracy measurements,
can be applied to refine these physical constants.
At present, however, the accuracy of the final value for the dissociation energy 
of the ground level of H$_2$ is restricted by the lack of the relativistic nuclear 
recoil contribution and the limited accuracy of the leading QED correction. 

\begin{table*}[!hbt]
\caption{Convergence of the lowest eigenvalue $E_{0,0}$ (in a.u.) and of the corresponding dissociation 
energy $D_{0,0}$ (in $\icm$) for H$_2$ with the basis set size $K$. 
Two-sector wave function was employed: 
(30,$\Omega$,19.19,$\beta^{(1)}$) and (30,$\Omega$-2,19.19,$\beta^{(2)}$).}
\label{Tconv}
\begin{tabular*}{\textwidth}{c@{\ }@{\extracolsep{\fill}}ccrw{4.17}w{7.12}}
\hline\hline\rule{0pt}{3ex}
$\Omega$ & $\beta^{(1)}$ & $\beta^{(2)}$ &  $K\quad$ & \cent{E_{0,0}\qquad} & \cent{D_{0,0}\qquad}  \\
\hline\\[-8pt]
      10 & 0.9304 & 2.664  &  36\,642 & -1.164\,025\,030\,822\,08    & 36\,118.797\,732\,723 \\
      11 & 0.953  & 3.041  &  53\,599 & -1.164\,025\,030\,870\,90    & 36\,118.797\,743\,437 \\
      12 & 0.978  & 3.45   &  76\,601 & -1.164\,025\,030\,880\,47    & 36\,118.797\,745\,538 \\
      13 & 1.011  & 3.20   & 106\,764 & -1.164\,025\,030\,882\,36    & 36\,118.797\,745\,953 \\
      14 & 1.039  & 2.80   & 146\,072 & -1.164\,025\,030\,882\,87    & 36\,118.797\,746\,064 \\
$\infty$ &        &        & $\infty$ & -1.164\,025\,030\,883\,1(3)  & 36\,118.797\,746\,10(3) \\
\hline\hline
\end{tabular*}
\end{table*}

\subsection{$J>0$ levels}

We consider here rotationally excited states of the ground 
vibrational level ($v=0$, $J=1$--$9$). In this case the admixture of states with
non-zero electronic angular momentum has to be taken into account in forming
the wave function (see Sec.~\ref{Sec:var} and~\ref{Sec:pert}). However, still a vast
contribution of energy comes from the $\Sigma$ wave function and the main effort
in the calculations has to be focused in the convergence of the energy within the space
formed by the $\Sigma$ basis functions. For this purpose, in analogy with the $J=0$ 
case described above, we composed the $\Sigma$ wave function of two sectors of basis functions
with a common $\alpha$ parameter: $(30,\Omega,\alpha,\beta^{(1)})$ and $(20,\Omega,\alpha,\beta^{(2)})$.
By increasing the shell parameter $\Omega$ we determined the extrapolated energy value
and its uncertainty. Sample data illustrating the energy convergence
for a selection of states are given in Table~\ref{TJconv}. A general observation
made from this table is that the convergence does not deteriorate significantly with the increasing
angular momentum $J$, so that for the highest state considered ($J=9$), the uncertainty
is about the same as that for $J=1$, amounting to $10^{-13}$ hartree.
In each case, the attained accuracy in dissociation energy is better than $5\cdot 10^{-8}\,\icm$.

\begin{table*}[!htb]
\caption{Convergence of the $\Sigma$-component of selected eigenvalues $E_{v,J}$ (in a.u.) with 
the increasing size of the basis set. Two-sector wave functions have been employed:
$(30,\Omega,\alpha,\beta^{(1)})$ and $(20,\Omega,\alpha,\beta^{(2)})$. 
$K$ is the total size of the basis set and $D_{v,J}$ --- the dissociation energy in $\icm$.}
\label{TJconv}
\begin{tabular*}{\textwidth}{r@{\ }@{\extracolsep{\fill}}cccrw{4.17}w{6.13}}
\hline\hline
\multicolumn{7}{c}{\rule{0pt}{3ex}$J=1$} \\
$\Omega$ & $\alpha$ & $\beta^{(1)}$ & $\beta^{(2)}$ & $K\quad$ & \cent{E_{0,1}\qquad} & \cent{D_{0,1}\qquad} \\
\hline\\[-8pt]
 9 & 16.92 & 0.866 & 2.183 &  27650 & -1.163\,485\,139\,541\,0    & 36\,000.305\,292\,83    \\
10 & 16.93 & 0.912 & 2.487 &  40950 & -1.163\,485\,139\,578\,5    & 36\,000.305\,301\,06    \\
11 & 16.99 & 0.929 & 3.064 &  61152 & -1.163\,485\,139\,584\,7    & 36\,000.305\,302\,43    \\
12 & 16.93 & 0.968 & 3.265 &  82600 & -1.163\,485\,139\,586\,3    & 36\,000.305\,302\,77    \\
13 & 16.93 & 1.05  & 3.4776& 117936 & -1.163\,485\,139\,586\,6    & 36\,000.305\,302\,83    \\
$\infty$ & &   &    & $\infty\quad$ & -1.163\,485\,139\,586\,7(1) & 36\,000.305\,302\,85(2) \\
\hline
\multicolumn{7}{c}{$J=5$} \\
$\Omega$ & $\alpha$ & $\beta^{(1)}$ & $\beta^{(2)}$ & $K\quad$ & \cent{E_{0,5}\qquad} & \cent{D_{0,5}\qquad} \\
\hline\\[-8pt]
 9 & 16.69 & 0.8599 & 2.282 &  28756 & -1.156\,095\,754\,663\,5    & 34\,378.522\,770\,78 \\
10 & 15.62 & 0.8998 & 2.659 &  42588 & -1.156\,095\,754\,701\,1    & 34\,378.522\,779\,04 \\
11 & 15.62 & 0.9225 & 3.026 &  61152 & -1.156\,095\,754\,707\,3    & 34\,378.522\,780\,40 \\
12 & 15.62 & 0.974  & 3.300 &  85904 & -1.156\,095\,754\,708\,8    & 34\,378.522\,780\,73 \\
13 & 15.62 & 1.013  & 3.350 & 117936 & -1.156\,095\,754\,709\,1    & 34\,378.522\,780\,79 \\
$\infty$ & &   &    & $\infty\quad$  & -1.156\,095\,754\,709\,2(1) & 34\,378.522\,780\,82(3) \\
\hline
\multicolumn{7}{c}{$J=9$} \\
$\Omega$ & $\alpha$ & $\beta^{(1)}$ & $\beta^{(2)}$ & $K\quad$ & \cent{E_{0,9}\qquad} & \cent{D_{0,9}\qquad} \\
\hline\\[-8pt]
 9 & 16.45 & 0.847 & 2.265 &  28756 & -1.141\,233\,218\,382\,6    & 31\,116.573\,099\,32 \\
10 & 16.29 & 0.888 & 2.641 &  42588 & -1.141\,233\,218\,421\,8    & 31\,116.573\,107\,91 \\
11 & 15.39 & 0.913 & 3.041 &  61152 & -1.141\,233\,218\,428\,5    & 31\,116.573\,109\,38 \\
12 & 15.39 & 0.955 & 2.750 &  85904 & -1.141\,233\,218\,429\,8    & 31\,116.573\,109\,66 \\
13 & 15.39 & 0.988 & 3.425 & 117936 & -1.141\,233\,218\,430\,1    & 31\,116.573\,109\,74 \\
$\infty$ & &   &   & $\infty\quad$  & -1.141\,233\,218\,430\,2(1) & 31\,116.573\,109\,76(3) \\
\hline\hline
\end{tabular*}
\end{table*}

To evaluate the perturbational corrections $E_\Pi^{(2)}$, $E_\Pi^{(4)}$, and $E_\Delta^{(4)}$
we employed one-sector wave functions of proper symmetry. This time, the sector formally
depends on a single $\alpha$ and three $\beta$ ($\beta^\Sigma$, $\beta^\Pi$, $\beta^\Delta$)
parameters. However, numerical
experiments have shown that the optimal $\beta^\Delta$ is very close to the
optimal $\beta^\Sigma$, and for convenience it was fixed at the value of the
latter, $\beta^\Delta=\beta^\Sigma$. Table~\ref{TNconv} contains sample results
of our convergence study of the three energy corrections computed according
to the formulas presented in Sec.~\ref{Sec:pert}. An inspection of the last three
columns of the table gives a view on the rate of convergence and the estimated
uncertainties of particular corrections. It also informs
how fast the particular corrections grow with increasing $J$.

We would like to emphasize that the perturbational approach described in Sec.~\ref{Sec:pert} is numerically, 
within the assumed goal of accuracy, totally equivalent to the variational one. 
The perturbational approach requires three decompositions of pertinent chunks 
($\mathbb{H}_{\Sigma\Sigma}$, $\mathbb{H}_{\Pi\Pi}$, and $\mathbb{H}_{\Delta\Delta}$)
of the Hamiltonian matrix, whereas in the variational approach the matrix must 
be decomposed as a whole. Because the decomposition effort is proportional 
to cubic size of the matrix ($\sim\!K^3$), the perturbational approach is, 
for large matrices, significantly more effective than the variational one.
We have confronted the results obtained for $D_{0,J}$ in both ways and obtained agreement 
better than $10^{-10}\,\icm$. This numerical agreement 
shows also that consideration of only those three corrections ($E_\Pi^{(2)},E_\Pi^{(4)},E_\Delta^{(4)}$)
is totally sufficient for our purposes. 

\begin{table*}[!htb]\footnotesize
\caption{Convergence of the $\Sigma$-, $\Pi$-, and $\Delta$-components of selected 
rotational energy levels (in a.u.) with the increasing size of the basis set. 
One-sector wave functions have been employed:
$(30,\Omega,\alpha,\beta^{\Sigma},\beta^{\Pi},\beta^{\Delta})$. 
$K$ is the total size of the basis set.}
\label{TNconv}
\begin{tabular*}{\textwidth}{r@{\extracolsep{\fill}}cccrw{3.12}w{5.8}w{4.7}w{4.7}}
\hline\hline\rule{0pt}{3ex}
$\Omega$ & $\alpha$ & $\beta^{\Sigma}=\beta^{\Delta}$ & $\beta^{\Pi}$ & $K\quad$ & 
\cent{E_\Sigma\qquad} & \cent{E_\Pi^{(2)}\cdot 10^8} & \cent{E_\Pi^{(4)}\cdot 10^{13}} & \cent{E_\Delta^{(4)}\cdot 10^{13}} \\
\hline
\multicolumn{9}{c}{$J=1$} \\
 9 & 19.84 & 0.930 & 0.875 &  37429 & -1.163\,485\,134\,761 & -3.272\,719\,00 & -0.492\,866 & 0.0 \\
10 & 19.82 & 0.973 & 0.904 &  55965 & -1.163\,485\,138\,021 & -3.272\,723\,62 & -0.492\,887 & 0.0 \\
11 & 19.83 & 1.008 & 0.940 &  81144 & -1.163\,485\,138\,961 & -3.272\,723\,88 & -0.492\,892 & 0.0 \\
12 & 19.84 & 1.049 & 0.968 & 114716 & -1.163\,485\,139\,347 & -3.272\,724\,31 & -0.492\,894 & 0.0 \\
$\infty$ & &   &    & $\infty\quad$ & -1.163\,485\,140\,3(3)& -3.272\,725(1)  & -0.492\,896(2)& 0.0 \\
\hline
\multicolumn{9}{c}{$J=5$} \\
 9 & 22.19 & 0.923 & 0.873 &  58669 & -1.156\,095\,749\,888 & -48.914\,586\,6 & -104.5339 &  -31.0603 \\
10 & 22.19 & 0.964 & 0.893 &  87342 & -1.156\,095\,753\,145 & -48.914\,652\,4 & -104.5382 &  -31.0614 \\
11 & 22.20 & 0.999 & 0.936 & 126120 & -1.156\,095\,754\,085 & -48.914\,655\,3 & -104.5390 &  -31.0616 \\
12 & 22.20 & 1.044 & 0.964 & 177644 & -1.156\,095\,754\,471 & -48.914\,660\,9 & -104.5395 &  -31.0618 \\
$\infty$ & &   &    & $\infty\quad$ & -1.156\,095\,754\,8(4)& -48.914\,67(1)  & -104.541(2)& -31.0622(4) \\
\hline
\multicolumn{9}{c}{$J=9$} \\
 9 & 23.90 & 0.915 & 0.860 &  58669 & -1.141\,233\,213\,617 & -145.750\,039 & -836.830  & -259.851 \\
10 & 23.91 & 0.958 & 0.889 &  87342 & -1.141\,233\,216\,875 & -145.750\,228 & -836.864  & -259.857 \\
11 & 23.91 & 0.994 & 0.926 & 126120 & -1.141\,233\,217\,812 & -145.750\,240 & -836.871  & -259.857 \\
12 & 23.92 & 1.034 & 0.953 & 177644 & -1.141\,233\,218\,194 & -145.750\,258 & -836.875  & -259.858 \\
$\infty$ & &   &    & $\infty\quad$ & -1.141\,233\,218\,6(4)& -145.750\,28(2)&-836.880(5)&-259.859(1) \\
\hline\hline
\end{tabular*}
\end{table*}

The final dissociation energies obtained for the lowest nine rotational levels $J=1,\dots,9$ 
of the ground vibrational state are presented in Table~\ref{TFconv}. 
The total energy was composed of the $E_\Sigma$ energy evaluated using a two-sector 
wave function and the subsequent perturbational corrections
$E_\Pi^{(2)},E_\Pi^{(4)},E_\Delta^{(4)}$ obtained from a one-sector wave function. 
It can be seen that the final accuracy is determined mainly by the accuracy achieved 
for the $E_\Sigma$ term. For higher $J$ though, the uncertainty originating from $E_\Pi^{(2)}$
becomes significant.
The second-order correction resulting from the coupling
of the nuclear rotational angular momentum with the electronic $\Pi$-state is indispensable
for accurate calculation, even for the $J=1$ level. This correction increases
with growing $J$ proportionally to $J(J+1)$ and for $J=9$ contributes to the dissociation
energy as much as $0.3\,\icm$. The fourth-order $\Pi$-states correction 
and the $\Delta$-states correction are much smaller but grow even more rapidly ($\sim\![J(J+1)]^2$) 
and become important when higher-$J$ states or still higher accuracy is of interest.

In Table~\ref{TFconv} we compared the total nonadiabatic dissociation energy with
the energy obtained from the second order nonadiabatic perturbation theory (NAPT)\cite{PK15}.
The difference between these two numbers comes from the higher order $\mathcal{O}(m_\mathrm{e}/M)^3$
terms not included in the perturbational calculations of Ref.~\citenum{PK15}.

\begin{table*}[!htb]
\caption{Nonadiabatic dissociation energy ($D_{0,J}$) of the lowest rotational energy levels 
of the ground vibrational state.
The total as well as the $\Sigma$-, $\Pi$-, and $\Delta$-components are given (in $\icm$).
For comparison, results from the second order nonadiabatic perturbation theory (NAPT)\cite{PK15} 
are also given. The difference Total-NAPT reflects the value of the higher order terms
missing in the NAPT calculations.
}
\label{TFconv}
\begin{tabular*}{\textwidth}{c@{\extracolsep{\fill}}*{5}{w{5.14}}}
\hline\hline\rule{0pt}{3ex}
Component     & \cent{J=1}              & \cent{J=2}              & \cent{J=3}              \\
$E_\Sigma$    & 36\,000.305\,302\,85(2) & 35\,764.407\,695\,23(2) & 35\,413.244\,980\,04(2) \\
$E_\Pi^{(2)}$ &       0.007\,182\,80    &       0.021\,536\,94    &       0.043\,040\,02    \\
$E_\Pi^{(4)}$ &       0.000\,000\,01    &       0.000\,000\,10    &       0.000\,000\,38    \\
$E_\Delta^{(4)}$&     0.000\,000\,00    &       0.000\,000\,02    &       0.000\,000\,10    \\
Total         & 36\,000.312\,485\,66(2) & 35\,764.429\,232\,28(2) & 35\,413.288\,020\,54(2) \\[0.5ex]
NAPT          & 36\,000.312\,413        & 35\,764.429\,157        & 35\,413.287\,941        \\[1ex]
\hline\\[0ex]
Component     & \cent{J=4}              & \cent{J=5}              & \cent{J=6}              \\
$E_\Sigma$    & 34\,949.943\,579\,00(2) & 34\,378.522\,780\,82(3) & 33\,703.780\,596\,09(3) \\
$E_\Pi^{(2)}$ &       0.071\,659\,75(1) &       0.107\,355\,29(1) &       0.150\,079\,08(2) \\
$E_\Pi^{(4)}$ &       0.000\,001\,04    &       0.000\,002\,29    &       0.000\,004\,39    \\
$E_\Delta^{(4)}$&     0.000\,000\,30    &       0.000\,000\,68    &       0.000\,001\,33    \\
Total         & 34\,950.015\,240\,09(2) & 34\,378.630\,139\,08(3) & 33\,703.930\,680\,89(4) \\[0.5ex]
NAPT          & 34\,950.015\,154        & 34\,378.630\,045        & 33\,703.930\,576        \\[1ex]
\hline\\[0ex]
Component     & \cent{J=7}              & \cent{J=8}              & \cent{J=9}              \\
$E_\Sigma$    & 32\,931.166\,238\,29(3) & 32\,066.646\,553\,99(3) & 31\,116.573\,109\,76(3) \\
$E_\Pi^{(2)}$ &       0.199\,778\,78(3) &       0.256\,399\,28(3) &       0.319\,884\,89(4) \\
$E_\Pi^{(4)}$ &       0.000\,007\,59    &       0.000\,012\,16    &       0.000\,018\,37    \\
$E_\Delta^{(4)}$&     0.000\,002\,33    &       0.000\,003\,76    &       0.000\,005\,70    \\
Total         & 32\,931.366\,026\,99(4) & 32\,066.902\,969\,19(4) & 31\,116.893\,018\,72(5) \\[0.5ex]
NAPT          & 32\,931.365\,910        & 32\,066.902\,838        & 31\,116.892\,871        \\[1ex]
\hline\hline
\end{tabular*}
\end{table*}

\section{Summary and outlook}

The numerical results presented in this work concern only the hydrogen molecule,
but the method described is applicable to any four-particle Coulomb system.
The achieved accuracy surpasses that available to date for such systems.
Until now, such an accuracy was available only for systems with three or fewer particles.
Apart from the high accuracy, the main advantage of this method is in
the formalism that enables straightforward calculations for non-zero rotational angular
momentum. This feature opens up a window for accurate prediction
of nonrelativistic energy for all bound levels in such systems.

The numerical results presented here constitute an introductory but indispensable part
of a larger project aimed at predicting the energy levels of H$_2$ with an accuracy of
$10^{-6}\,\icm$. This part must be followed by accurate (at least 1 ppm) calculations of
the leading relativistic ($\sim\alpha^2$) and QED ($\sim\alpha^3$) corrections 
as well as corrections resulting from higher-order ($\sim\alpha^4$ and $\alpha^5$) 
contributions and other tiny effects like the finite size of the nucleus 
or {\em gerade-ungerade} mixing\cite{PK11}.
We have recently evaluated the relativistic correction 
and the higher-order QED corrections in the Born-Oppenheimer regime.\cite{PKP17,PKCP16}
The new nonadiabatic wave functions will enable us to take into account also
the finite nuclear mass effects in the corrections mentioned above.

\section*{Acknowledgements}

This research was supported by NCN Grants No.
2012/04/A/ST2/00105 (K.P.) and 2017/25/B/ST4/01024 (J.K.), 
as well as by a computing grant from the
Poznan Supercomputing and Networking Center, and by
PL-Grid Infrastructure.


\bibliography{naH2} 
\bibliographystyle{apsrev}

\end{document}